\documentclass[twocolumn]{aastex61}

\usepackage{graphicx}
\usepackage{xcolor}
\usepackage[sort&compress]{natbib}
\usepackage[hang,flushmargin]{footmisc}
\usepackage[counterclockwise]{rotating}
\usepackage{amsmath,natbib,graphicx}
\usepackage{lipsum}
\usepackage{gensymb}

\shortauthors{Salazar, et al.}
\shorttitle{The effect of substellar continent size on the ocean dynamics of Proxima Centauri b}
\usepackage{enumitem,xcolor}

\usepackage{lineno}

\begin{document}
\graphicspath{ {./} }
\DeclareGraphicsExtensions{.pdf,.eps,.png}

\title{The effect of substellar continent size on ocean dynamics of Proxima Centauri b}

\author{Andrea M. Salazar}
\affiliation{Department of the Geophysical Sciences, University of
  Chicago, 5734 South Ellis Avenue, Chicago, IL 60637} 
  
  \author{Stephanie L. Olson}
\affiliation{Department of the Geophysical Sciences, University of
  Chicago, 5734 South Ellis Avenue, Chicago, IL 60637}

\author{Thaddeus D. Komacek}
\affiliation{Department of the Geophysical Sciences, University of
  Chicago, 5734 South Ellis Avenue, Chicago, IL 60637}
  
 \author{Haynes Stephens}
\affiliation{Department of the Geophysical Sciences, University of
  Chicago, 5734 South Ellis Avenue, Chicago, IL 60637}
  
\author{Dorian S. Abbot}
\affiliation{Department of the Geophysical Sciences, University of
  Chicago, 5734 South Ellis Avenue, Chicago, IL 60637}

\correspondingauthor{Andrea M. Salazar}
\email{amjsalazar@uchicago.edu}

\begin{abstract}

The potential habitability of tidally locked planets orbiting M-dwarf stars has been widely investigated in recent work, typically with a non-dynamic ocean and without continents. On Earth, ocean dynamics are a primary means of heat and nutrient distribution. Continents are a critical source of nutrients, strongly influence ocean dynamics, and participate in climate regulation. In this work, we investigate how the size of a substellar land mass affects the ocean’s ability to transport heat and upwell nutrients on the tidally locked planet Proxima Centauri b using the ROCKE-3D coupled ocean-atmosphere General Circulation Model (GCM). We find that dayside ice-free ocean and nutrient delivery to the mixed layer via upwelling are maintained across all continent sizes. We also find that Proxima Centauri b’s climate is more sensitive to differences among atmospheric GCMs than to the inclusion of ocean dynamics in ROCKE-3D. Finally, we find that Proxima Centauri b transitions from a “lobster” state where ocean heat transport distributes heat away from the substellar point to an “eyeball” state where heat transport is restricted and surface temperature decreases symmetrically from the substellar point when the continent size exceeds $\sim$20\% of the surface area. Our work suggests that both a dynamic ocean and continents are unlikely to decrease the habitability prospects of nearby tidally locked targets like Proxima Centauri b that could be investigated with future observations by the James Webb Space Telescope (JWST).

\end{abstract}

\section{Introduction}

M-dwarf stars comprise nearly 80\% of all stars in the galaxy and about one out of six M-dwarfs have an Earth-sized planet within their Habitable Zone \citep{dressing2015occurrence}. The Habitable Zone (HZ) is defined as the range of distances from the star in which a planet can have liquid water at its surface   \citep{kasting1993habitable}. M-dwarfs stars are smaller and dimmer than the Sun, and consequently their HZs are much closer to the star, which makes a HZ planet likely to become tidally locked in a 1:1 spin orbit state in which one side of the planet is in constant daytime while the other is in perpetual night. Many planets in the HZ of M-dwarfs have already been observed \citep{anglada2016terrestrial,gillon2017seven,luger2017seven}. The discovery of Proxima Centauri b was especially exciting, because it orbits an M-dwarf star only 4.2 lightyears away and is very likely to be tidally locked \citep{anglada2016terrestrial}. Its proximity to Earth will make it a prime target for future observations with the James Webb Space Telescope (JWST).

These exciting discoveries have spurred theoretical research into the habitability of tidally locked planets like Proxima Centauri b. These habitability studies have mostly used Global Climate Models (GCMs) that assume either no continents \citep{hu2014role,fujii2017nir,checlair2019no,yang2019ocean}, a static “slab” ocean \citep{joshi1997simulations,leconte20133d,yang2013stabilizing,yang2014strong,checlair2017no}, or both \citep{merlis2010atmospheric,wang2014climate,turbet2016habitability,boutle2017exploring,haqq2018demarcating,komacek2019atmospheric,quintana2020first}. These studies tend to agree that without a dynamic ocean, the climate of a tidally locked planet will settle into an “eyeball” state  \citep{pierrehumbert2010palette} characterized by roughly radially symmetric surface temperatures that decrease away from the substellar point.

On Earth, ocean dynamics play a vital role in heat and nutrient distribution. The importance of ocean circulation on Earth has inspired a few recent studies of ocean circulation on extrasolar planets  \citep{cullum2014importance,hu2014role,yang2014water,checlair2019no,jansen2019climates,olson2019oceanographic,yang2019ocean}. These have established that on tidally locked plants, ocean dynamics play a key role in heat redistribution from the dayside to the nightside.  Moreover, a dynamic ocean introduces asymmetries in the surface temperature distribution that have been described as a “lobster” state because of the way the ocean transports heat from the substellar point towards the nightside of the planet \citep{hu2014role}. This greatly expands the area of open ocean and increases global average temperatures. 

Additionally, ocean dynamics are an important control on the replenishment of essential nutrients to the mixed layer that are lost when biomass sinks to the bottom of the ocean (the so-called biological pump, \citep{marshall2008atmosphere}). The mixed layer is the area of the surface ocean that has nearly uniform properties because it is well mixed. The depth of the mixed layer varies spatially and is determined by the vertical potential density gradient, which is influenced by the sea surface temperature and salinity, as well as wind strength near the surface. Upwelling into the mixed layer is generally discouraged by stable density stratification, with warm sunlit surface water sitting atop cold deep water. Upwelling occurs in regions where ocean currents diverge such that conservation of mass requires deep water to move upward into the mixed layer. Ocean dynamics therefore exert direct control on the distribution of nutrients in the surface ocean, and therefore photosynthetic life on Earth.  In an exoplanet context, \citet{olson2019oceanographic} explored oceanographic constraints on life, such as areas of upwelling. They argued that planets with higher rates of upwelling may be more favorable for photosynthetic life. 

Continental configuration is an important control on both ocean currents and climate. The importance of continents for ocean circulation and climate can be seen in idealized GCM simulations where only slivers of continents can disrupt ocean flows and lead to major reorganizations in climate \citep[e.g.,][]{ferreira2011climate,rose2013role,marshall2014ocean}. Moreover, the carbonate-silicate cycle requires precipitation on exposed continents to regulate atmospheric CO$_2$ \citep{walker1981negative}. Continental weathering also delivers nutrients to the ocean that are essential for oceanic life, potentially making continents necessary not only for long-term climate stability, but also for life. On tidally locked planets, we expect continents to impose similar constraints on ocean currents and climate. In the context of tidally locked planets, \citet{wieczorek2007gravity} suggested that true polar wander will align topographical anomalies with the planet-star axis, meaning that continents will tend to congregate at the substellar or anti-stellar point \citep{leconte2018continuous}.

\citet{lewis2018influence} used the Met Office Unified Model (UM) to explore how substellar continents affect the climate of Proxima Centauri b with no ocean dynamics. They found that the habitability of Proxima Centauri b is largely insensitive to the size of the substellar continent. For continents ranging from 4\% to 39\% of the total surface area of the planet, a patch of above-freezing temperatures persisted at the substellar point and temperatures elsewhere never fell below 125 K, avoiding condensation of CO$_2$ onto the surface. Increasing continent size also increased temperature contrasts between the day and the nightside and tended to cool the planet on a global average. 

In this work, we use the ROCKE-3D coupled ocean-atmosphere General Circulation Model (GCM) to investigate how substellar continent size on Proxima Centauri b affects two vital controls on habitability: ocean heat transport and nutrient upwelling. In Section \ref{ROCKE}, we describe our model setup with ROCKE-3D, including assumptions of planetary parameters of Proxima Centauri b. In Section 3, we 
investigate the effect of substellar continent size on ocean heat transport and nutrient upwelling on tidally locked planets. We discuss the effect of ocean dynamics on the habitability of Proxima Centauri b and the potential observability of biosignatures in Section \ref{discussion} and conclude in Section 5.

\section{Methods} \label{ROCKE}

\begin{figure*}
\begin{center}
  \includegraphics[width=1\textwidth]{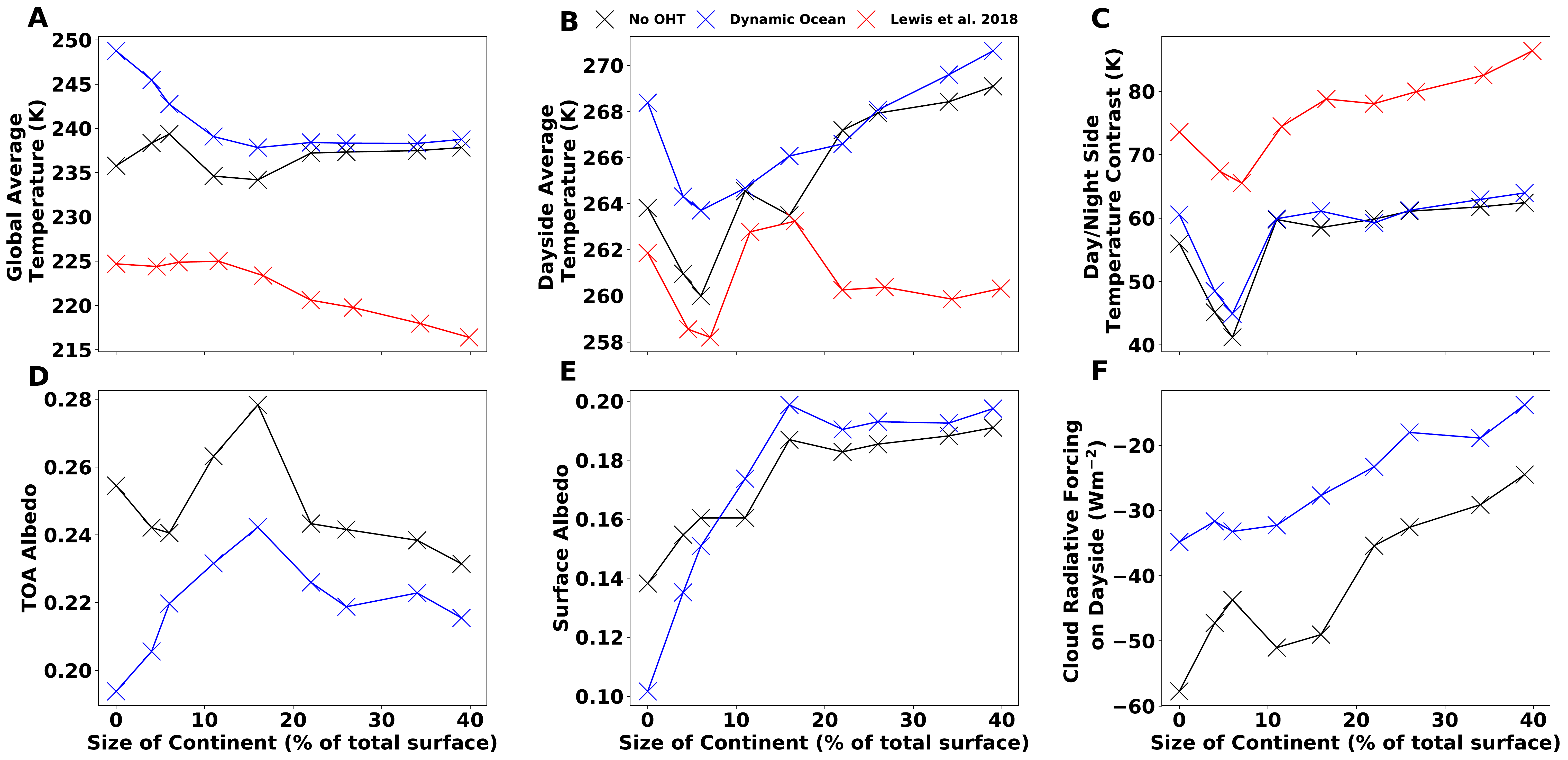}
  \caption{\textbf{The effect of the dynamic ocean diminishes rapidly as continent fraction increases due to a reduction in ocean heat transport.} Once continent coverage reaches about 20\%, the no OHT and dynamic ocean simulations yield very similar results. Differences between this work and \cite{lewis2018influence} are due primarily to differences in prescribed albedo of land and ice-free ocean as well as convection and cloud parameterizations. In every case, our results are more sensitive to differences between ROCKE-3D and UM than they are to the inclusion of ocean dynamics.}
  \label{fig:globtemp}
\end{center}
\end{figure*}

We use ROCKE-3D \citep{way2017resolving}, a coupled ocean-atmosphere 3D GCM adapted from the Goddard Institute for Space Studies (GISS) ModelE2. We use a 4\degree x 5\degree   \  latitude-longitude resolution with 40 vertical layers in the atmosphere and a topmost layer pressure of 0.1 mb. ROCKE-3D uses the radiative transfer model SOCRATES, described in detail by \citet{walters2019met}. SOCRATES was developed for the UM and was also used by \citet{lewis2018influence}. The stellar spectrum for Proxima Centauri was prepared by \citet{meadows2017proxima}, with $T_{eff} = 3042$ K and $R = 9.8\times 10^{7}$ m. 

We explore two ocean regimes to isolate the effect of ocean dynamics. The first is a slab ocean with constant 24 m mixed layer depth and lateral ocean heat transport set to zero (hereafter, no OHT), following \citet{lewis2018influence}. The second is a fully coupled dynamic ocean with a 5 layer resolution and a prescribed depth of 150 m. We use a relatively shallow ocean to reduce computational time. In both cases, sea surface salinity is 34.7 psu. Ocean, land, sea ice, and snow albedo prescriptions are described in detail by \citet{way2017resolving}. We ensure that our simulations reach convergence in global mean top-of-atmosphere net radiative flux.

Following \citet{lewis2018influence}, we model our planet after Proxima Centauri b with an assumed surface gravity, $g = 10.98$ m/s$^2$, and planetary radius, $r_p = 7127$ km. The planet receives an incident stellar flux of 881.7 W/m$^2$. We assume that Proxima Centauri b is tidally locked with an orbital period of 11.19 Earth days, and has zero eccentricity and obliquity. We assume a surface pressure of 1 bar and an N$_2$-dominated atmosphere with 280 ppm of CO$_2$. The substellar point is located at the equator at 0\degree  \   longitude. Continents are centered at the substellar point, and have sizes ranging from 0\% (aqua planet) to 39\% of the total surface area of the planet.

Following \citet{olson2019oceanographic}, we calculate upwelling as the vertical velocity at the base of the mixed layer. The mixed layer is the area of the ocean that has nearly uniform properties and is in equilibrium with the surface. ROCKE-3D defines the mixed layer depth (MLD) as the depth at which the potential density is 0.03 kg/m$^3$ higher than the surface \citep{large1994oceanic}. We further classify upwelling as coastal upwelling if any of the adjacent cells are land. We are especially interested in dayside upwelling since photosynthetic life requires access to sunlight.

\begin{figure*}
\begin{center}
  \includegraphics[width=\textwidth]{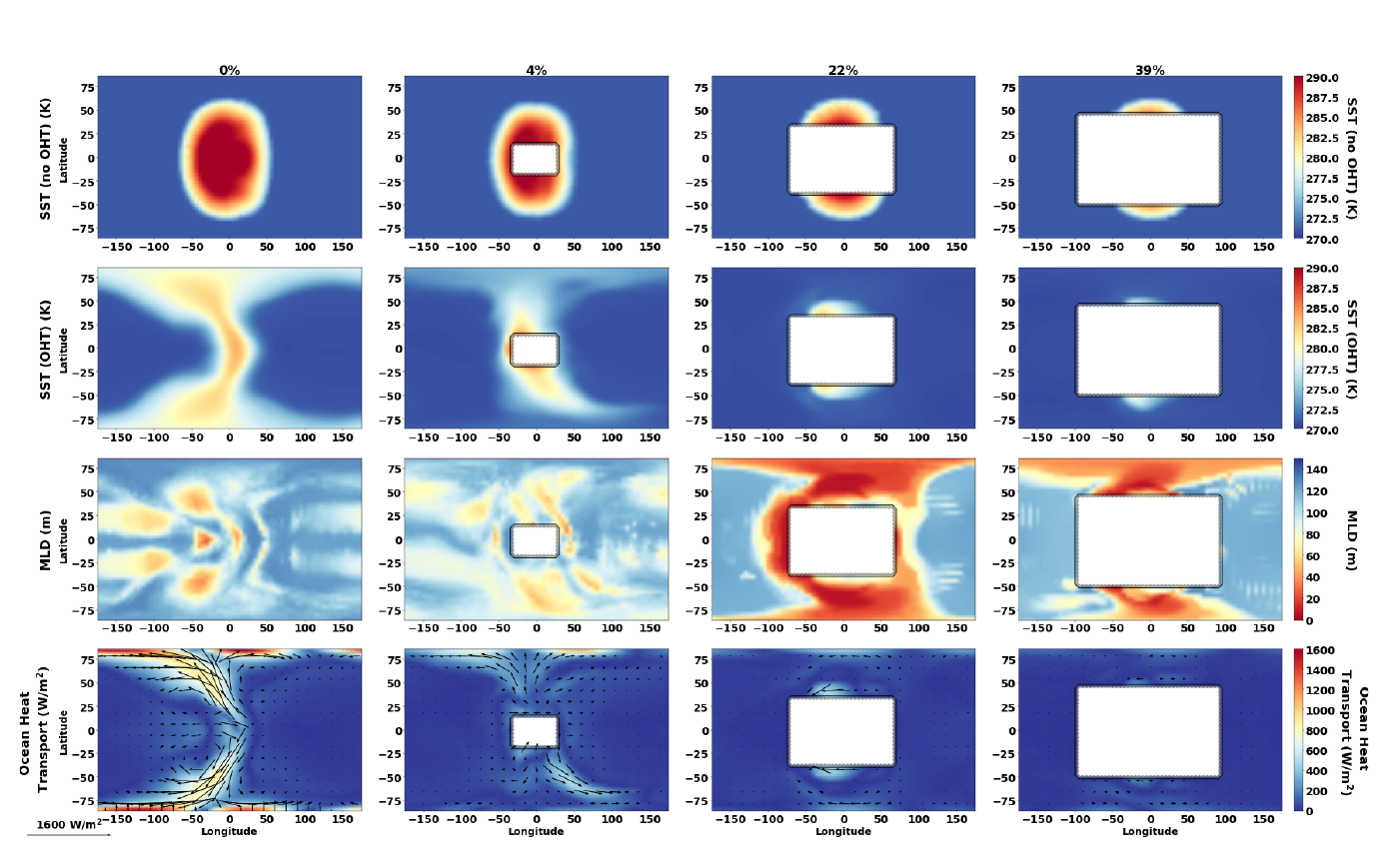}
  \caption{\textbf{Continents at the substellar point inhibit ocean heat transport, limiting the effect of ocean dynamics.} As continent size increases, the planet transitions from a ``lobster state" to an ``eyeball state", characteristic of a tidally locked planet with no ocean dynamics. The center of each map is the substellar point, and white boxes represent continents. The colorbars correspond to the entire row of plots. Arrows in the bottom row are normalized to 1600 Wm$^{-2}$ and are determined by the direction and magnitude of N-S and E-W heat flux. The colorbar in the bottom row represents the magnitude of ocean heat transport.}
  \label{fig:cmap}
\end{center}
\end{figure*}

\section{Results}

\subsection{Temperature Offset Between ROCKE-3D and UM} \label{lewisvsalazar}

We begin by comparing simulations in ROCKE-3D with a slab ocean (no OHT) to similar simulations in UM by \citet{lewis2018influence}. Figure \ref{fig:globtemp} shows how surface temperatures are affected by substellar continent size in both the dynamic and no OHT cases. The no OHT ROCKE-3D simulations have global mean surface temperatures 9-21 K higher than the UM simulations (Fig. 1(a)). This difference grows with increasing continent sizes, which is likely due to differences in prescribed land albedo. The UM used by \citet{lewis2018influence} prescribes an ice-free ocean albedo of 0.07 and a land albedo of 0.4. The prescribed albedos in ROCKE-3D are systematically lower, with an ice-free ocean albedo of 0.03 and a bare soil albedo of 0.2. For simulations with substellar continents, the large difference in land albedo means that less incident solar radiation is reflected, which leads to higher average temperatures. 

For the aqua planet simulation, the relatively small ice-free ocean albedo differences alone cannot account for the temperature differences between ROCKE-3D and UM. Additionally, ROCKE-3D and UM use the same radiative transfer model, SOCRATES, so radiative transfer calculations cannot account for the difference in temperature. We suspect that differences in the cloud and convection schemes, which can change the global-mean surface temperature by tens of Kelvin \citep{sergeev2020atmospheric}, are responsible for the differences between ROCKE-3D and UM when there are no continents.

\begin{figure*}
\begin{center}
  \includegraphics[width=1\textwidth]{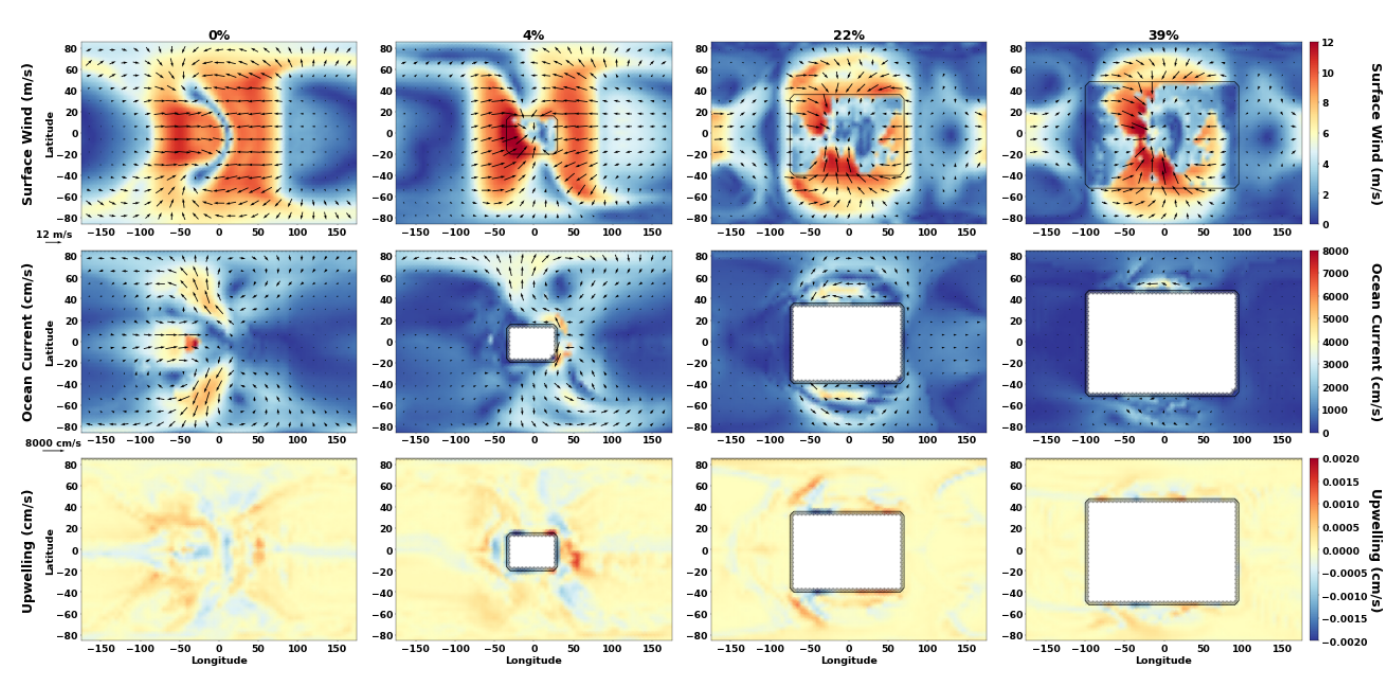}
  \caption{\textbf{Areas of upwelling are correlated with areas of divergent ocean currents, especially off the coast of continents.} Upwelling on the dayside is maintained for every continent size considered in this work. The vectors in the first and second row are determined from the direction and magnitude of N-S and E-W surface winds and ocean currents. The colorbars in the first two rows represent the magnitude of surface winds and ocean currents. Positive values of upwelling corresponds to upward movements of water at the base of the mixed layer while negative values correspond to downward movements.}
  \label{fig:upwell_map}
\end{center}
\end{figure*}

\subsection{Effect of Ocean Dynamics on Temperature} \label{dynontemp}

Next, we consider simulations with a fully dynamic ocean to investigate whether the inclusion of a dynamic ocean affects the habitability of tidally locked planets. From Figure \ref{fig:globtemp}, we see that the effect of including a dynamic ocean diminishes as substellar continent size increases. This is because a primary effect of ocean dynamics on a tidally locked planet is to transport heat away from the substellar point to the nightside. When a substellar continent of increasing size blocks heat redistribution, this effect weakens. Moreover, it is important to note that our results are much more sensitive to differences between ROCKE-3D and UM than they are to the inclusion of ocean dynamics. 

There is a small difference between the surface albedo in the dynamic ocean and no-OHT cases (Figure \ref{fig:globtemp}(e)). This is due to differences in sea ice coverage, which is impacted by ocean heat transport.

\subsection{Effect of a Substellar Continent on Ocean Dynamics} \label{lobstereyeball}

Previous work has shown that the insolation pattern on a tidally locked planet creates an ``eyeball” state in which surface temperatures decrease roughly symmetrically from the substellar point \citep{pierrehumbert2010palette}. In the first row of Figure \ref{fig:cmap}, we show sea surface temperatures in the no OHT case. Heat is concentrated near the substellar point, creating an ``eyeball" pattern. 

When ocean dynamics are included, the ocean transports heat from the substellar point, which expands the area of deglaciated ocean, forming a ``lobster state” \citep{hu2014role}. The 0\% land-coverage case in the second row of Figure \ref{fig:cmap} demonstrates how a dynamic ocean transports heat towards the nightside such that the sea surface temperature pattern has ``lobster claws'' of warm ocean away from the substellar point. In general, global-mean temperature is higher and the day/night temperature contrast is smaller in the lobster state due to the transport of heat away from the substellar point. 

Adding a continent at the substellar point disrupts the ocean's ability to transport heat. Figure \ref{fig:cmap} demonstrates the planet’s transition from the lobster state with efficient ocean heat transport to the eyeball state with diminished ocean heat transport as continent size grows. Once the continent covers about 20\% of the surface of the planet, dynamic ocean simulations yield results fairly similar to corresponding no OHT simulations.

The bottom row of Figure \ref{fig:cmap} demonstrates the effect of continent size on ocean heat transport in the dynamic ocean simulations. On the aqua planet (0\% continent size),  OHT efficiently transports heat from the substellar point towards the nightside, creating the lobster pattern in the sea surface temperature map. However, even the addition of a small, 4\% continent dramatically disrupts OHT. Though the continent never entirely shuts off OHT, a 22\% continent restricts OHT enough that the planet settles into an eyeball state similar to the no OHT case.

\subsection{Effect of a Substellar Continent on Nutrient Fluxes} \label{upwelling}

\begin{figure*}
\begin{center}
  \includegraphics[width=\textwidth]{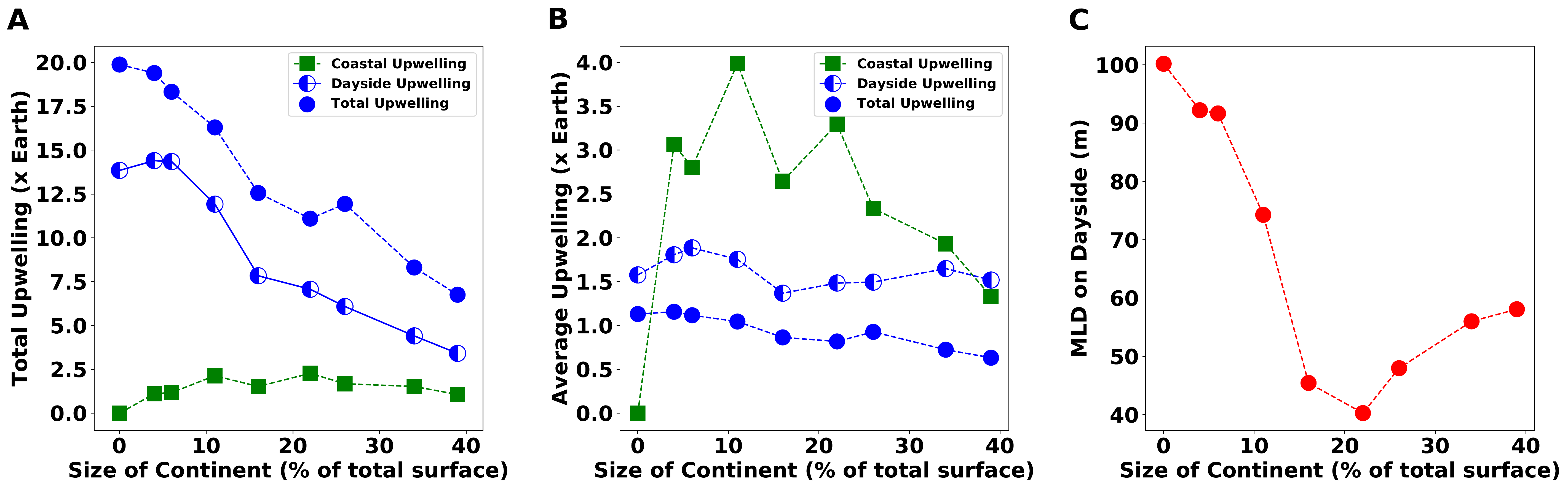}
  \caption{\textbf{Upwelling on the dayside alone is significantly higher than total upwelling in an Earth-like simulation.} Larger continents have more coast line, allowing for more total coastal upwelling. The mixed layer shallows as a result of the lobster to eyeball transition and deepens again as the substellar continent covers the warmest areas of the ocean. Figures 4(a) and 4(b) are normalized to an Earth-like simulation performed by \citet{olson2019oceanographic}. In Figure 4(b), coastal upwelling is averaged over the length of coastline, dayside upwelling is averaged over ocean area on the dayside, and total upwelling is averaged over total ocean area.}
  \label{fig:upwell_MLD}
\end{center}
\end{figure*}

Figure \ref{fig:upwell_map} shows how divergent ocean currents determine areas of upwelling. In the third row of Figure \ref{fig:upwell_map}, upwelling occurs on the dayside for every continent size considered in this work. Upwelling corresponds to areas of divergent surface ocean currents shown in the second row of Figure \ref{fig:upwell_map}. This is because the conservation of mass requires an upwelling of water in areas of diverging surface currents. This upwelling may supply nutrients to the shallow ocean where photosynthesis is viable.

Figure \ref{fig:upwell_MLD} shows the change in upwelling and average mixed layer depth on the dayside of the planet as substellar continent size increases. Here, upwelling is defined as the area-weighted sum of vertical velocity at the base of the mixed layer. All values are normalized to the Earth-like baseline from \citet{olson2019oceanographic}.

Both total and dayside upwelling decrease with increasing continent size due to a general decrease in ocean area (Figure \ref{fig:upwell_MLD}(a)). Dayside upwelling constitutes the majority of total upwelling, which is favorable for photosynthetic life. The total (Figure \ref{fig:upwell_MLD}(a)) and average (Figure \ref{fig:upwell_MLD}(b)) dayside upwelling are significantly larger than the corresponding upwelling in an Earth-like simulation for all continent sizes.

The mixed layer depth (MLD) is the depth of the surface ocean that is homogenized by surface winds. On Earth, the MLD is sensitive to sea surface temperatures, which affect ocean stratification, and shallows in the summer when sea surface temperatures rise. In our simulations, the MLD is deepest in the aqua planet case and rapidly shallows with increasing substellar continent size, corresponding to changes in sea surface temperature and therefore ocean density stratification. As before, we can attribute this behavior to the transition from lobster to eyeball seen in Figure \ref{fig:cmap}. When ocean heat transport is very efficient, heat does not concentrate as strongly on the dayside, resulting in a deeper mixed layer. However, as ocean heat transport is inhibited by a continent, heat is trapped on the dayside, shallowing the mixed layer there. With continents larger than about 20\%, the behavior of the average MLD on the dayside is dominated by the effect of decreasing ocean area on the dayside. As the continent grows, it does so at the expense of the warmest portions of the ocean with the shallowest MLD, so the average MLD is deeper.

\section{Discussion} \label{discussion}

Ocean dynamics play a key role in heat transport and in previous work have been shown to impact the total habitable area on tidally locked planets. In this work, we showed that the habitability of Proxima Centauri b with substellar continents is not very sensitive to the inclusion of a dynamic ocean. Differences in surface albedo, cloud, and convective parameterizations between ROCKE-3D and UM have a much larger effect on climate than ocean dynamics. Differences across climate models are a huge uncertainty. This is consistent with the relatively large effect that seemingly small differences in modeling assumptions can have when GCMs are applied to exoplanets \citep{yang2016differences,yang2019simulations}. 

In every model scenario we considered, the dayside of the planet had liquid water, satisfying the most basic definition of habitability. Next, we investigated other controls on photosynthetic life, like nutrient availability and access to sunlight. The vast majority of oceanic life on Earth resides in the mixed layer. On Earth, the light penetration depth (the depth at which solar radiation can penetrate the water) is of the order of 100 m. Photosynthetic life is more productive when the mixed layer is shallow compared to the light penetration depth. Since tidally locked planets orbit red-dwarf stars, the incident solar radiation is shifted towards the infrared, which has a shallower light penetration depth on the order of 10 m \citep{kaltenegger2019dark}. Therefore, we can expect that life on tidally locked planets would favor areas of the ocean with very shallow mixed layers. 

We found that the MLD initially shallows with increasing continent size due to a decrease in the efficiency of OHT. Additionally, continents provide a source of coastal upwelling, which delivers nutrients to the mixed layer. Continents are also a source of nutrients like phosphorus from chemical weathering that are vital for photosynthesis \citep{glaser2020detectability}. However, if a continent is large enough to cover more than about half of the dayside, this begins to deepen the average mixed layer depth on the dayside by covering ocean that would have been extremely warm. This would force life to exist closer to the nightside of the planet, which could restrict photosynthetic life. 

Photosynthetic life requires both light and nutrients. We showed that a majority of upwelling, which provides nutrients, occurs on the dayside, where there is light. Photosynthesis is thus viable on synchronously rotating planets, including  planets with substellar land mass. A high concentration of photosynthetic life can affect the spectral appearance of a planet, and thus might be remotely detectable \citep{olson2019oceanographic, schwieterman2018exoplanet}. Moreover, photosynthetic life, which translates stellar energy into chemical energy in the form of thermodynamic disequilibrium may be uniquely detectable with JWST, even if it does not produce O$_2$ \citep{krissansen2018detectability,krissansen2018disequilibrium}.

In this work we prescribed a 150 m ocean to reduce computational time. There is evidence that the depth of an ocean can influence surface temperatures \citep{del2019habitable,yang2019ocean}. Additionally, the MLD in simulations with small continents nearly reached 150 m in some areas, so the shallowness of the ocean used in this work may impart some uncertainty to our results. However, day side mixed layers are typically 40-100 m deep, rarely approaching the depth of the ocean except when large continents force photosynthetic marine life to the margins of the day side (Figure \ref{fig:upwell_MLD}).  Future work could examine how increasing ocean depth affects our results, but we expect that our interpretations regarding the suitability of tidally locked planets for photosynthetic life are robust against this uncertainty.

\section{Conclusions}
In this work, we used ROCKE-3D to investigate how substellar continents on Proxima Centauri b affect two vital ocean processes: ocean heat transport and nutrient upwelling. Our main conclusions are as follows:
\begin{enumerate}
\item The habitability of Proxima Centauri b is largely independent of ocean dynamics for the planetary parameters we considered. In both the dynamic and slab ocean scenario, liquid water is found on the dayside regardless of substellar continent size. 
\item Surface temperature results are more sensitive to differences in the surface albedo, convection, and cloud schemes between ROCKE-3D and the Met Office Unified Model than the inclusion of ocean dynamics. 
\item Substellar continents reduce the efficiency of ocean heat transport. Our simulations suggest that Proxima Centauri b should transition from a lobster to an eyeball state as substellar continent size increases and inhibits ocean heat transport.
\item Proxima Centauri b experiences widespread dayside upwelling across all continent sizes. This is vital for returning nutrients to the surface ocean that are lost when biomass sinks to the deep ocean.
\item The average mixed layer depth on the dayside is lowest for continents between 16-26\% of the total surface area of the planet. Shallow mixed layers are preferable for photosynthetic life on planets around M-dwarfs due to shallower light penetration depths of their oceans. 

\end{enumerate}

For all of the continent sizes we considered, dayside surface ocean conditions are conducive to photosynthetic marine life. This is an important result given the prevailing view that continental weathering is essential for not only longterm climate stability, but also for life. This work suggests that tidally locked planets with surface oceans can maintain habitable conditions for life regardless of substellar continent size, which may further the case for extraterrestrial life on synchronously rotating worlds.

\acknowledgements
This work was completed with resources provided by the University of
Chicago Research Computing Center. This work was partially supported
by the NASA Astrobiology Program grant No. 80NSSC18K0829 and benefited
from participation in the NASA Nexus for Exoplanet Systems Science
research coordination network. S.L.O. acknowledges support from the
T.C. Chamberlin Post-doctoral Fellowship in the Department of
Geophysical Sciences at the University of Chicago. T.D.K. acknowledges
funding from the 51 Pegasi b Fellowship in Planetary Astronomy
sponsored by the Heising-Simons Foundation. We thank the referee for insightful comments that improved the manuscript.

\newpage

\bibliography{proxb_dynocean}


 

\end{document}